\documentclass{article}

    \PassOptionsToPackage{numbers, compress}{natbib}
  \usepackage[dblblindworkshop, final]{neurips_2026}
\usepackage{subcaption}
\usepackage{graphicx}
\usepackage{placeins}

\usepackage[utf8]{inputenc} % allow utf-8 input
\usepackage[T1]{fontenc}    % use 8-bit T1 fonts
\usepackage{hyperref}       % hyperlinks
\usepackage{url}            % simple URL typesetting
\usepackage{booktabs}       % professional-quality tables
\usepackage{amsfonts}       % blackboard math symbols
\usepackage{nicefrac}       % compact symbols for 1/2, etc.
\usepackage{microtype}      % microtypography
\usepackage{xcolor}         % colors

\usepackage{booktabs} % For \toprule, \midrule, \bottomrule
\usepackage{amsmath}  % For \text{} inside math mode

\title{End-to-end Particle Reconstruction with Transformers in High-Granularity Calorimeters}

\author{ 
  \bfseries Qiming Bao$^{*}$,
  Haotian Bi$^{*}$,
  Benedikt Maier,
  Alexander Tapper,
  Lauri Laatu
  \\[1.0em]
  \begin{tabular}{c}
    Imperial College London, United Kingdom \\
    $^{*}$\textit{Equal contribution}
  \end{tabular}
}

\begin{document}

\maketitle

\begin{abstract}
%  The HGCAL of the CMS experiment will provide detailed measurements of particle showers after the HL-LHC upgrade, but it will also present a challenging reconstruction problem when several showers overlap. This project studies how detector hits can be reconstructed into a set of particles with 3 different machine learning models using a calorimeter dataset inspired by CMS HGCAL data. The results show that modern generative and query methods provide effective alternatives for particle reconstruction.

The High-Luminosity upgrade of the Large Hadron Collider (HL-LHC) will drastically increase the number of simultaneously interacting particles, posing a severe challenge to current particle reconstruction methods. While next-generation highly granular calorimeters offer unprecedented 3D measurement capabilities, reconstructing particles from heavily overlapping energy showers remains computationally difficult. This study investigates how raw detector hits can be accurately reconstructed using end-to-end transformer based methods on a simulated calorimeter dataset. We evaluate and compare three distinct machine learning frameworks, focusing on two generative architectures: diffusion and flow matching as well as query-based set prediction. Our results demonstrate that these advanced methods provide highly effective and scalable alternatives for resolving complex overlapping particle showers in conditions encountered in future calorimeters.

\end{abstract}

\section{Introduction}

% This section describes the problem, existing methods and our approach
% Example of data, model architectures if there is space for that
%Deep generative models has developed rapidly in recent years, with new architectures designed for a growing range of tasks. Adapting ideas from modern architectures may provide another way to approach the problems found in Physics research. 
%We introduce this possibility in the context of particle reconstruction for the High Granularity Calorimeter (HGCAL) of the Compact Muon Solenoid (CMS) experiment at CERN. HGCAL is being developed as part of the High-Luminosity Large Hadron Collider (HL-LHC) upgrade and will measure particle showers with fine spatial and longitudinal granularity \cite{cmsHGCAL}. The detector response is represented as sparse and unordered 3D hits. The task is to identify the particles that produced these hits and reconstruct their properties. This is challenging because particle showers can overlap, many collisions can happen at the same time and the number of particles is unknown. 
% The purpose of the clustering algorithm when applied to shower reconstruction is to group together individual energy deposits (hits) originating from a particle shower.

\begin{figure}[htbp]
  \centering
  % First subfigure
  \begin{subfigure}{0.39\textwidth}
    \centering
    \includegraphics[width=\linewidth]{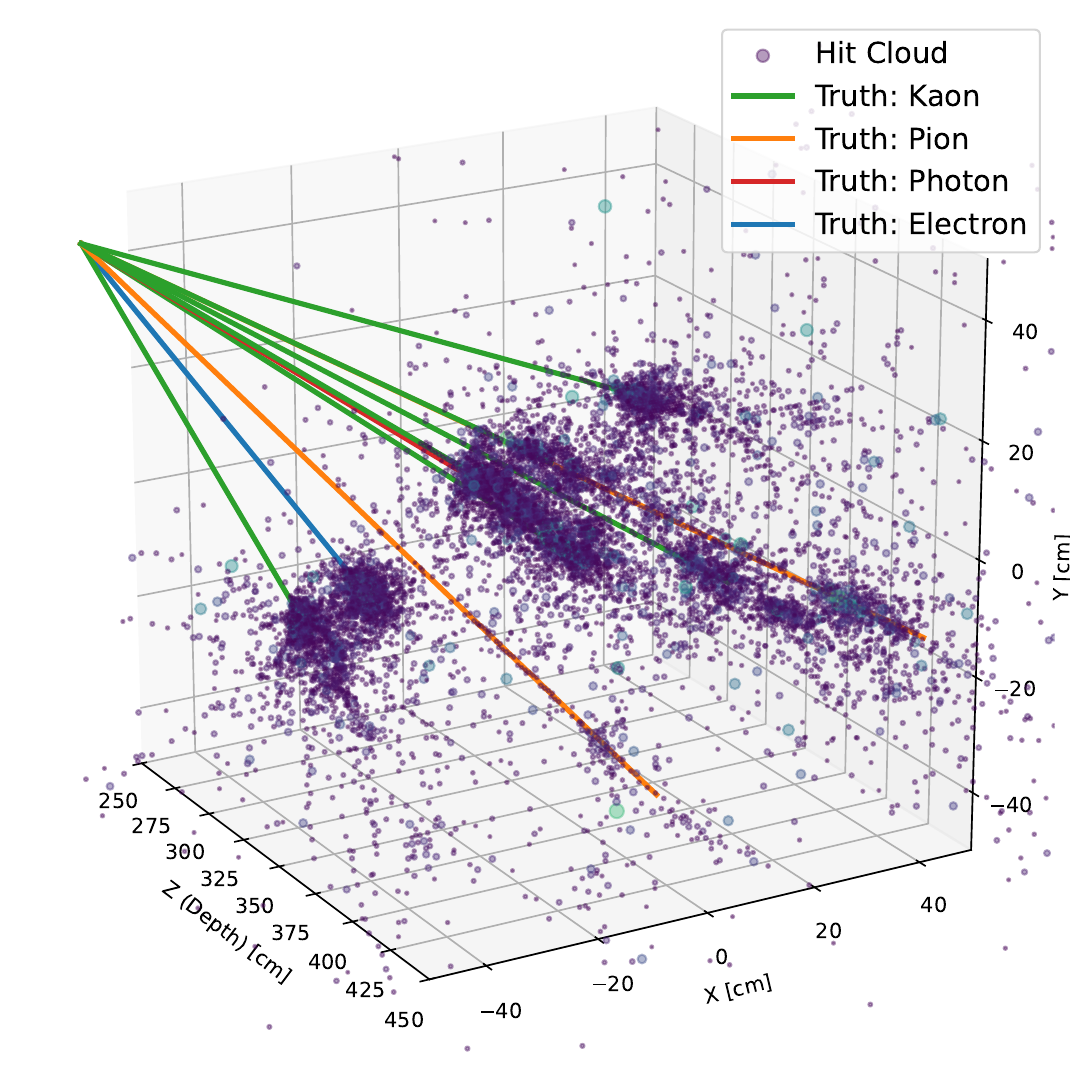}
    \caption{Calorimeter showers of 10 particles.}
    \label{fig:deposit}
  \end{subfigure}\hfill
  % Second subfigure
  \begin{subfigure}{0.6\textwidth}
    \centering
    \includegraphics[width=\linewidth]{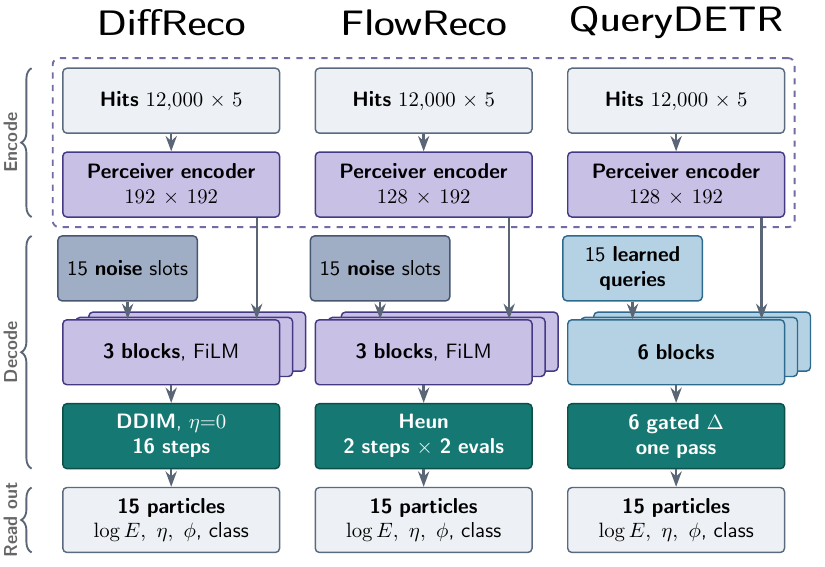}
    \caption{Model architectures.}
    \label{fig:architectures}
  \end{subfigure}\hfill
  % Third subfigure
  %\begin{subfigure}{0.32\textwidth}
  %  \centering
  %  \includegraphics[width=\linewidth]{images/i3.pdf}
  %  %\caption{Third stage}
  %  \label{fig:sink3}
  %\end{subfigure}
  
  %\caption{asd}
  \label{fig:description}
\end{figure}

As machine learning continues to advance rapidly, adapting cutting-edge architectures offers promising new pathways to solve complex data challenges in high-energy physics. We explore this potential in the context of particle reconstruction for the High Granularity Calorimeter (HGCAL)~\cite{cmsHGCAL}, currently under construction for the Compact Muon Solenoid (CMS)~\cite{CMS} experiment at CERN. Designed for the High-Luminosity Large Hadron Collider (HL-LHC) upgrade, the HGCAL will be able to record particle showers with 48 detecting layers per endcap and contains a total of around 5 million independent detection channels. The detector output is represented as a point cloud of sparse, unordered 3D hits. Reconstructing the identities and properties of the particles that generated these hits is highly non-trivial: the exact number of incident particles is unknown, numerous collisions occur simultaneously, and their resulting energy showers heavily overlap. An example of overlapping particles showers can be seen in figure~\ref{fig:deposit}. We introduce new machine learning based methods that reconstruct particles directly from hits and are able to effectively deal with overlapping particle showers in cases where traditional clustering methods fail. Previous work on neural network based reconstruction has been done in~\cite{qasimE2E,contrastiveCalo,clusTEX,gnnOverlap,juGNN,hgcalGNN, riemannianDiffusion, riemannianFlow}.

\section{Methods}

We present three transformer~\cite{attention} based end-to-end methods for particle reconstruction with similar parameter count, the architectures can be seen in figure~\ref{fig:architectures}. All models' input are individual calorimeter hits of position x, y, z (converted to $\eta$, $ sin(\phi) $, $ cos(\phi) $, z) and energy while the output contains up to 15 particles with energy, position (represented as $\eta$ and $\phi$), and class (particle ID).
The dataset contains 100,000 Geant4 events \cite{geant4} where the cell energies are true Geant4 energy deposits integrated per cell corresponding to an ideal detector response. Each original event contains 1--10 particles with energies of 20--200~GeV, emitted within a cone of 1$^\circ$ half-angle so that they hit a local region of radius $\approx$21~cm ($\approx$0.14~m$^2$) on the detector front face. The events therefore represent local regions of the detector rather than full HL-LHC events: with 10 particles the local particle density ($\approx$70~m$^{-2}$) is comparable to that expected from 200 pile-up interactions at $|\eta| \approx 1.5$--2, whereas the total detector occupancy is far below that of a full HL-LHC event and the particles are considerably more energetic than typical pile-up particles. The full dataset uses train-val-test split of 80\%-10\%-10\%. After filtering out events that contain only muons, these sets contain 78,047, 9,742 and 9,749 events. The neural networks learn to reconstruct electrons, photons, pions and kaons while ignoring the deposits by muons.

\subsection{QueryDETR}

QueryDETR adapts the Detection Transformer \cite{detr} to directly predict particle states from a set of learned queries. Up to 12 000 hits are first reduced to a 128 long representation using a Perciever~\cite{perceiver} encoder. Each query consists of a latent representation that stores learned features and a reference state that contains initial energy and directional coordinates. These queries pass through 6 blocks where each block contains both self-attention which allows them to interact and preventing duplicate reconstructions, and cross-attention which extracts relevant shower data from the calorimeter hits. After gathering this information, each block iteratively refines the energy and trajectory of the particle candidates. During training, a Hungarian matching \cite{kuhn} pairs the predictions of the model with the true particles.
%, using assignments from the final block to supervise the earlier stages.

\subsection{DiffReco}

DiffReco treats particle reconstruction as a denoising process inspired by object detection diffusion model DiffusionDet \cite{diffusionDet}. During training, the model organizes events into 15 candidate slots comprising both real particles and background, which are then corrupted with different levels of Gaussian noise based on a cosine schedule \cite{ddpm}. The neural network learns to reverse this corruption by predicting a target that combines the clean particle state and the added noise state \cite{salimansHo}, using encoded detector memory and noise-level data integrated through Feature-wise Linear Modulation \cite{film} to adapt to different stages of noisiness. During inference, the model starts with completely noisy candidates and employs an iterative sampling procedure to progressively denoise step-by-step until the clean particle states are fully recovered \cite{ddim}. Once the denoising is complete, output heads evaluate each candidate to determine whether it is a real particle with energy, position and class or should not be reconstructed.

\subsection{FlowReco}

FlowReco uses flow matching \cite{flowMatching, flowMatchingGuide} to transform initial random noise candidates into real particles. During the training phase, the model learns a continuous and straight-line path between these random starting points and their assigned true targets by predicting the flow velocity or vector field, which consists of the direction and rate of change, at various intermediate steps, all guided by Perciever encoded representation. When applied during inference, where the actual targets are unknown, the model begins with random noise and gradually evolves the particle candidates coordinates by solving an ordinary differential equation (ODE). To do this accurately, it employs Heun's method \cite{heunODE}, a 2-step solver that evaluates the network's predicted direction at the current position, estimates the candidate's next position, and evaluates the direction at this provisional state again, averaging the 2 predictions to precisely guide the candidates to their final reconstructed states. The classification head works the same way as DiffReco.

\subsection{CLUEstering}
CLUEstering is the baseline reconstruction model used in this project. It is based on CLUstering by Energy (CLUE) algorithm, which was developed for the fast clustering of hits in high-granularity calorimeters \cite{clue, cluesteringSoftware}. CLUE begins by calculating a local energy density for every calorimeter hit. With the local density and the nearest-higher distance, the hits will be divided by 3 groups: cluster seeds with a high density and far from other high density hits, outliers with low density and far from a denser region and followers which are the remaining hits. In this work the input for CLUEstering is the 3D position of the hits x, y, z and energy. For more details see Ref. \cite{clue}.

\section{Results}

\begin{figure}[htbp]
  \centering

  \begin{subfigure}[t]{0.49\textwidth}
    \centering
    \includegraphics[width=\linewidth]{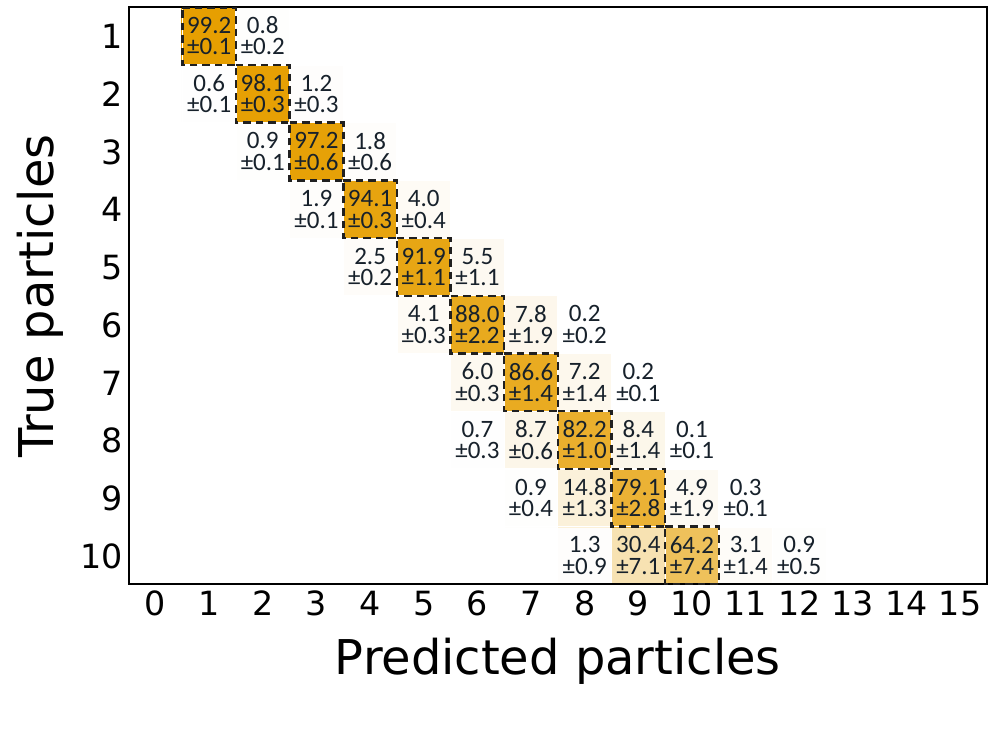}
    \caption{QueryDETR.}
    \label{fig:confusion1}
  \end{subfigure}\hfill%
  \begin{subfigure}[t]{0.49\textwidth}
    \centering
    \includegraphics[width=\linewidth]{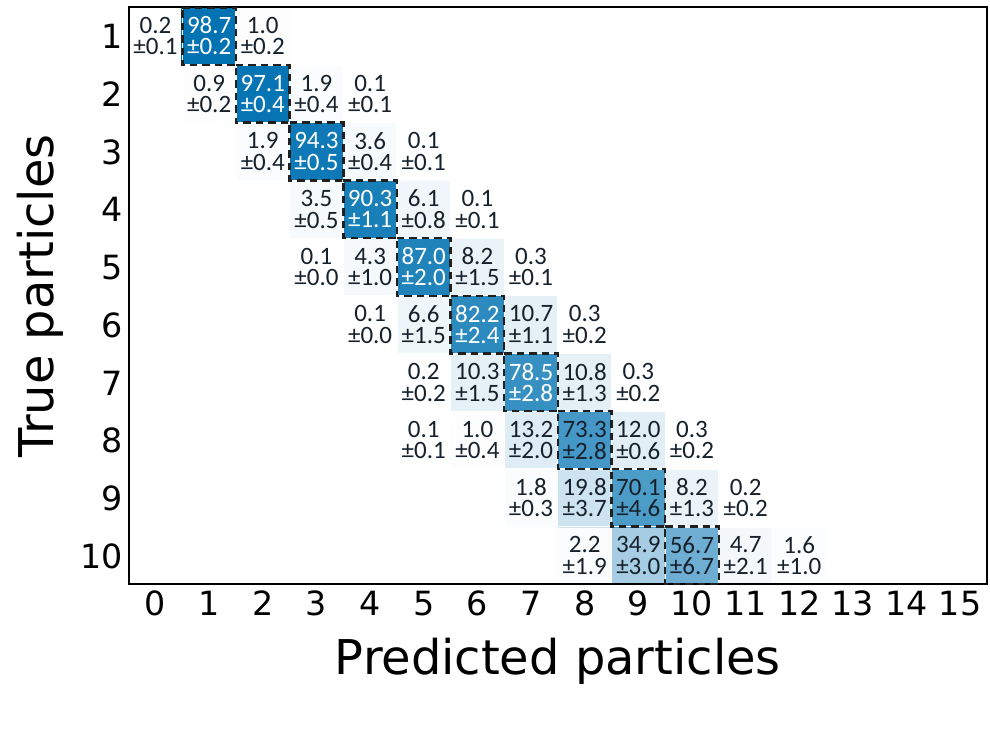}
    \caption{DiffReco.}
    \label{fig:confusion2}
  \end{subfigure}\hfill%
  \begin{subfigure}[t]{0.49\textwidth}
    \centering
    \includegraphics[width=\linewidth]{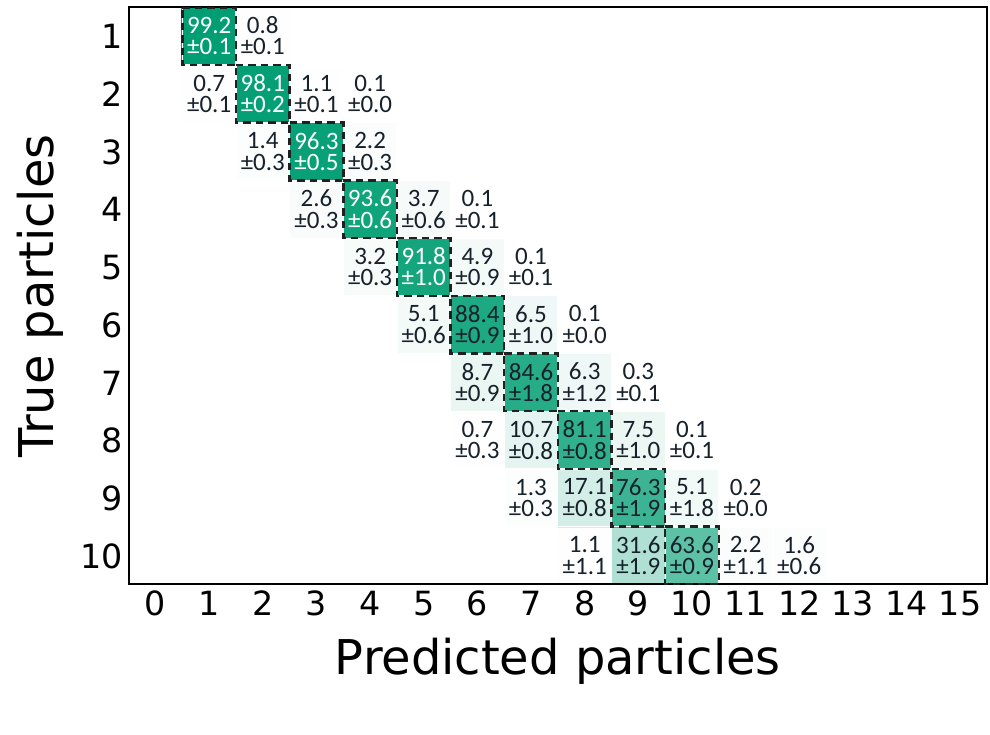}
    \caption{FlowReco.}
    \label{fig:confusion3}
  \end{subfigure}\hfill%
  \begin{subfigure}[t]{0.49\textwidth}
    \centering
    \includegraphics[width=\linewidth]{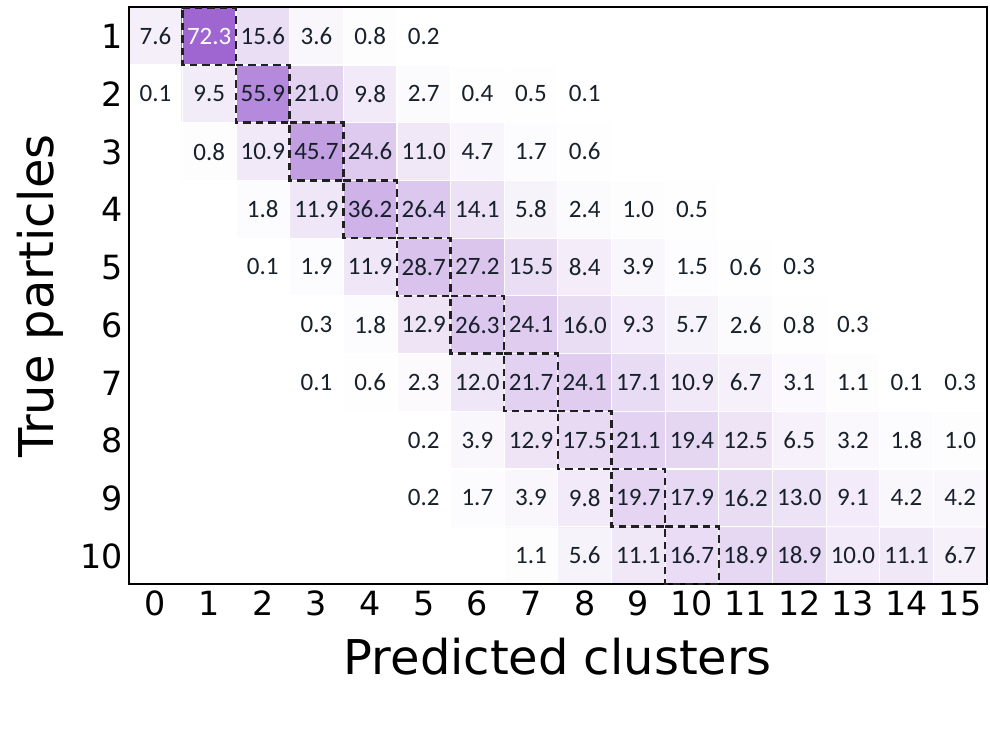}
    \caption{CLUEstering.}
    \label{fig:confusion4}
  \end{subfigure}

  \caption{
    Particle-multiplicity matrices for (a) QueryDETR, (b) DiffReco, (c) FlowReco and (d) CLUEstering. All valid predictions are counted, including unmatched and duplicate predictions. Rows are normalized to 100\%, and dashed boxes mark equal predicted and true counts. Neural network entries show the mean $\pm$ standard deviation across five seeds. QueryDETR has the strongest diagonal in most rows. CLUEstering shows a broader distribution.
  }
  \label{fig:particle_multiplicity}
\end{figure}

Figure~\ref{fig:particle_multiplicity} compares the true and predicted particle multiplicities for QueryDETR, DiffReco, FlowReco and the
CLUEstering baseline. A successfully reconstructed particle must be within $\Delta R \leq 0.1$ computed as distance in the coordinate space $ \sqrt{\Delta \eta ^2 + \Delta \phi ^2} $. All three neural network architectures exhibit a strong concentration along the diagonal (equal predicted and true particle counts), significantly outperforming CLUEstering.

%To isolate successful reconstructions, the learned models only count predictions matched to a truth particle within a spatial distance of $\Delta R \leq 0.1$. Consequently, these distributions reflect accurately recovered particles rather than total raw predictions.
%DiffReco achieves the highest exact-recovery rate for events with 1 to 8 truth particles; for instance, it successfully recovers every particle in 83.5\% of 6-particle events, compared to 80.6\% for QueryDETR and 79.7\% for FlowReco. In contrast, CLUEstering produces a much broader distribution with frequent over- and under-counting, dropping from a 70.4\% perfect recovery rate in single-particle events to just 15.3\% in 10-particle events.

% PREVIOUS VERSION: Figure~\ref{fig:Energy_residual} presents the energy resolution, $E_{\mathrm{pred}}-E_{\mathrm{true}}$ for the correctly reconstructed particles. The neural networks all demonstrate a sharp peak near zero. FlowReco yields the narrowest residual distribution with a standard deviation of 13.71~GeV, followed closely by DiffReco at 13.83~GeV and QueryDETR at 14.79~GeV. The CLUEstering baseline exhibits substantially broader tails, showing a pronounced tendency to underestimate the particle energies.

\begin{figure}[htbp]
  \centering

  \begin{subfigure}[t]{0.33\textwidth}
    \centering
    \includegraphics[width=\linewidth]{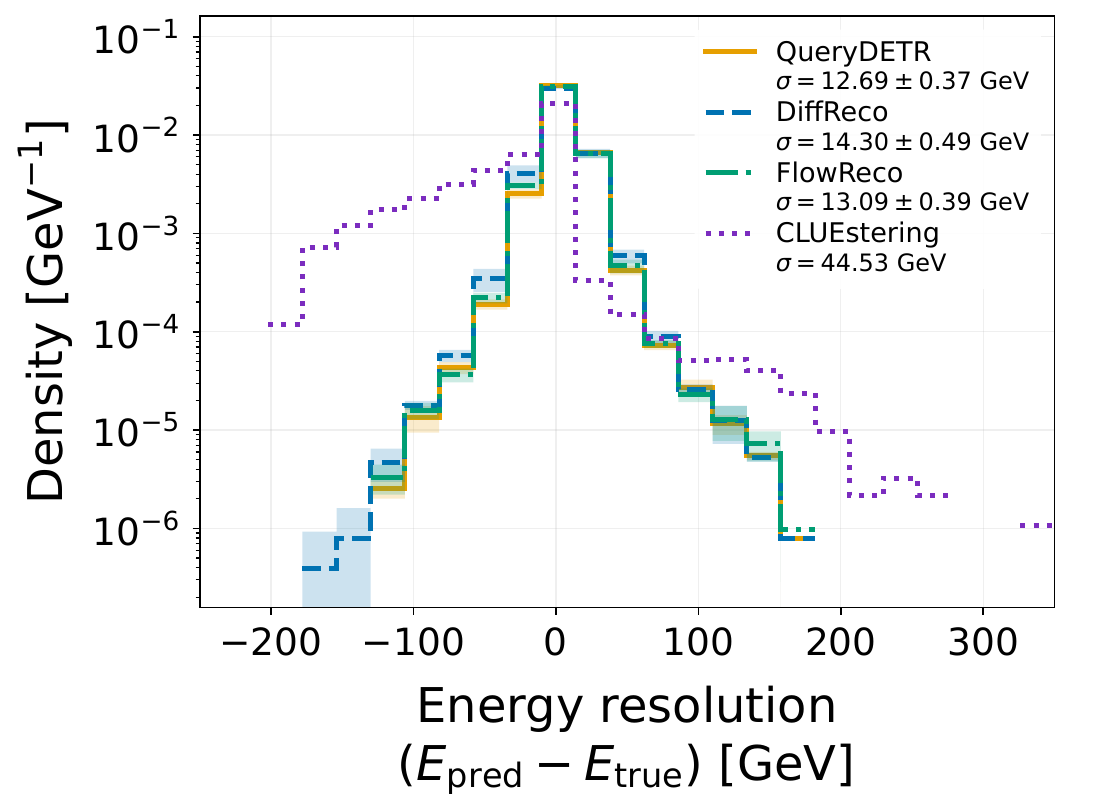}
    \caption{Energy resolution for reconstructed particles.}
    \label{fig:Energy_residual}
  \end{subfigure}\hfill
  \begin{subfigure}[t]{0.33\textwidth}
    \centering
    \includegraphics[width=\linewidth]{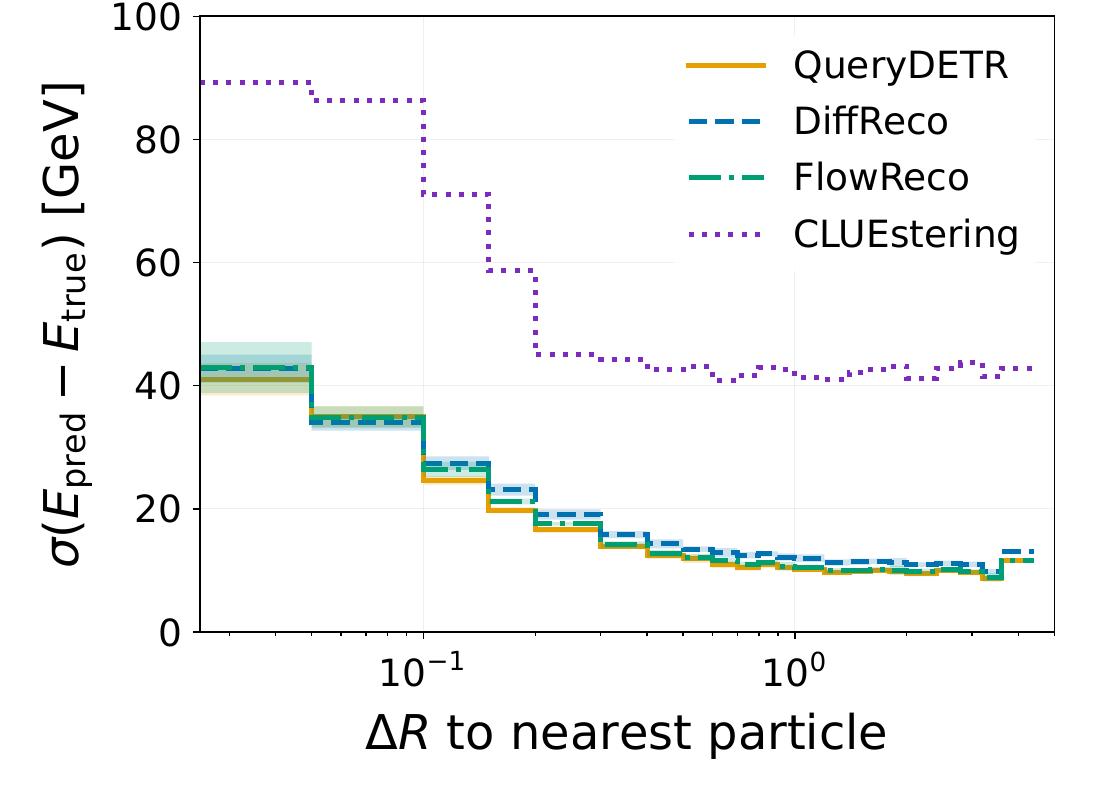}
    \caption{Energy resolution vs proximity.}
    \label{fig:Energy_resolution_vs_proximity}
  \end{subfigure}
  \begin{subfigure}[t]{0.33\textwidth}
    \centering
    \includegraphics[width=\linewidth]{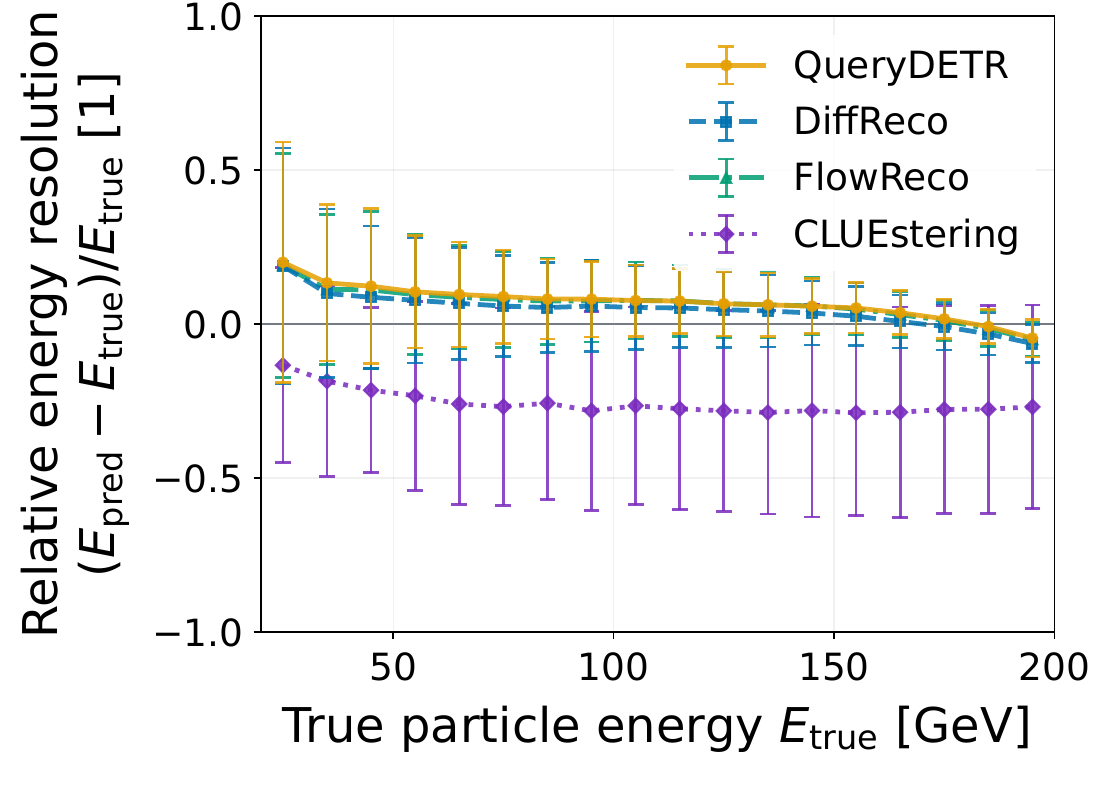}
    \caption{Relative error as a function of true particle energy.}
    \label{fig:rel_e}
  \end{subfigure}
  
  \caption{
    Energy reconstruction for particles matched within $\Delta R \leq 0.1$: (a) normalized energy-residual distributions; (b) residual standard deviation versus distance to the nearest other truth particle; and (c) mean relative error in 10\,GeV true-energy bins. Neural network error bars in (a,b) show the mean and standard deviation across five seeds. Panel (b) uses events with at least two true particles. Each neural network in (c) uses its seed with the lowest test-set energy-resolution standard deviation. Bars in (c) show the standard deviation of relative energy resolution in each bin. QueryDETR has the narrowest distribution in (a). All methods have poorer energy resolution for nearby particles.
  }
  \label{fig:results}
\end{figure}

Figure~\ref{fig:Energy_residual} presents the energy resolution, $E_{\mathrm{pred}}-E_{\mathrm{true}}$, for correctly reconstructed
particles. The neural networks all demonstrate a sharp peak near zero, with residual standard deviations of $12.69\pm0.37$~GeV for QueryDETR, $14.30\pm0.49$~GeV for DiffReco and $13.09\pm0.39$~GeV for FlowReco. These results show similar energy-reconstruction performance, with QueryDETR giving the smallest width. CLUEstering has a substantially broader distribution, with a standard deviation of 44.53~GeV and a pronounced negative tail, indicating underestimated particle energies. 

% Fix "Accepted matches" should be changed to where accepted matches is done: energy resolution (something else?). Explain R95 in caption, add standard deviation in the tables between different seeds. Fix MAE: Should be computed as mean of every event's abs(pred-true)/true particles

% PREVIOUS VERSION: The effect of overlapping showers is shown in figure~\ref{fig:Energy_resolution_vs_proximity} with energy resolution for reconstructed particles as a function of distance ($\Delta R$) to the nearest particle. As expected, energy reconstruction precision degrades as particles become more collimated.
% At spatial separations below 0.05, the energy residual widths of the neural models span 41--47~GeV due to the inherent difficulty of resolving heavily overlapping showers. This precision improves considerably as showers separate, tightening to widths of 11--12~GeV at distances around 1.0. Across the full range, the machine learning methods have similar performance with both DiffReco and FlowReco generally performing slighly better than QueryDETR and all three greatly outperform CLUEstering.

The effect of overlapping showers is shown in Figure~\ref{fig:Energy_resolution_vs_proximity}. The energy-residual width is plotted against the angular distance to the nearest other truth particle. For the closest particles, the neural network standard deviations are 41--43~GeV. They decrease to around 9--12~GeV at separations near $\Delta R=1$.  All three models follow the same general trend: energy assignment becomes more precise as the showers become more separated. CLUEstering has broader residuals throughout the displayed range.

% Possible UPDATE: Figure 3.C)

Figure~\ref{fig:rel_e} shows the relative energy resolution, $(E_{\mathrm{pred}}-E_{\mathrm{true}})/E_{\mathrm{true}}$, as a function of true particle energy. The neural networks tend to overestimate the energy at the lower end of the range, with this bias decreasing as the energy increases and becoming slightly negative near the upper end. CLUEstering shows a larger negative bias across the range.

%This comparison shows why both the spread and the energy dependence of the residuals are important when evaluating reconstruction.

%Across the populated separation range, FlowReco generally maintains the most constrained resolution widths, while CLUEstering remains consistently broader regardless of particle proximity.

% PREVIOUS VERSION: Table~\ref{tab:results_comparison} summarizes the results for both accuracy and throughput. DiffReco misses the smallest fraction of truth particles, at 3.54\%, followed by FlowReco at 4.43\% and QueryDETR at 5.13\%. It also gives the lowest normalized particle-count error and the narrowest angular error distribution, while FlowReco gives the narrowest accepted energy-residual distribution. These results agree with the trends in Figures~\ref{fig:particle_multiplicity} and~\ref{fig:results}. The best throughput for the machine learning methods is achieved by FlowReco due to the effectiveness of flow matching which gives good accuracy at only 2 refinement steps which results to 86\% of the speed of CLUEstering.

Table~\ref{tab:results_comparison} brings together particle recovery, classification, energy and directional reconstruction, computational
cost and throughput. The three neural networks recover 96\% of truth particles, compared with 87.20\% for CLUEstering. For particle recovery we only consider fully missed particles as not recovered and duplicates in the same region are not penalized.
QueryDETR and FlowReco have nearly identical recovery rates, while, QueryDETR gives the smallest energy-residual width and the lowest
$R_{95}(\Delta R)$ among accepted matches. Combined accuracy groups the predictions into electromagnetic, hadronic and background categories gives combined accuracies above 97\% for all three models since the data does not include any tracking information. 
% HERE describe the inference setup, batch sizes etc

Inference throughput is measured on an A100-SXM4-80GB, timing the full test set from normalized hits in host RAM to final particle outputs. Neural network inference uses JAX just-in-time (JIT) compiled models (compile time is excluded from the timing) with pipelined data transfer and mixed precision, using FP16 for dense and attention operations and FP32 for softmax and layer normalization. The selected neural network batch sizes are 256 for QueryDETR and 512 for DiffReco and FlowReco. CLUEstering uses C++/CUDA implementation at batch size 1024.
Both exceed the measured CLUEstering throughput of 10145 events per second. QueryDETR and FlowReco also achieve similar throughput speeds for the inference, processing 12725 and 12042 events per second, respectively. DiffReco processes 7455 events per second due to the required high number of 16 denoising updates. FlowReco obtains its reconstruction results using two Heun updates, each requiring two network evaluations. These measurements show that the number of repeated evaluations is important for inference cost, alongside the model architecture and execution configuration.

% give the batch numbers here too, but not in the caption!

% Possible UPDATE: STILL WAITING FOR GFLOPS AND SPEEED. Table~\ref{tab:results_comparison} summarizes the results for both performance and throughput. All three models have a close particle recovery. QueryDETR gives the lowest angular error distribution and the narrowest accepted energy-residual distribution. These results agree with the trends in Figures~\ref{fig:particle_multiplicity} and~\ref{fig:results}. The best throughput for the machine learning methods is achieved by FlowReco and QueryDETR due to the effectiveness of flow matching which gives good accuracy at only 2 refinement steps and QueryDETR _______ which results to XXXXX\% of the speed of CLUEstering.

%\vspace{2cm}

\begin{table*}[htbp]
    \centering
    \caption{Computational cost, reconstruction performance and throughput. Particle recovery is the fraction of truth particles matched within $\Delta R \leq 0.1$. Combined accuracy groups predictions into electromagnetic, hadronic and background, including correctly unused background slots and penalising missed and duplicate predictions. $\sigma(\Delta E)$ is the standard deviation of energy resolution among accepted matches; $R_{95}(\Delta R)$ is the 95th percentile of their angular distances. Neural network results with uncertainties show mean $\pm$ standard deviation across five seeds.}
  \label{tab:results_comparison}
    \small
    \renewcommand{\arraystretch}{1.15}
    \resizebox{\textwidth}{!}{%
    \begin{tabular}{l c c c c c c c c}
        \toprule
        \textbf{Method} &
        \textbf{GFLOP/ev} &
        \textbf{Steps} &
        \textbf{Params.} &
        \shortstack{\textbf{Particle}\\\textbf{Recovery} [\%]} &
        \shortstack{\textbf{Combined}\\\textbf{Accuracy} [\%]} &
        $\boldsymbol{\sigma(\Delta E)}$ [GeV] &
        $\boldsymbol{R_{95}(\Delta R)}$ &
        \shortstack{\textbf{Throughput}\\\textbf{(ev/s)}} \\
        \midrule
        CLUEstering & n/e & n/a & n/a & 87.20\% & n/a & 44.53 & 0.0544 & 10145 \\
        QueryDETR & \textbf{4.620} & n/a & 4.71\,M & \textbf{96.57 $\pm$ 0.19\%} & 97.75 $\pm$ 0.12\% & \textbf{12.69 $\pm$ 0.37} & \textbf{0.0505 $\pm$ 0.0009} & \textbf{12725} \\
        DiffReco & 6.256 & 16 & 3.30\,M & 95.92 $\pm$ 0.62\% & 97.33 $\pm$ 0.36\% & 14.30 $\pm$ 0.49 & 0.0553 $\pm$ 0.0028 & 7455 \\
        FlowReco & 4.727 & 2 $\times$ 2 & \textbf{3.29\,M} & 96.55 $\pm$ 0.18\% & \textbf{97.81 $\pm$ 0.15\%} & 13.09 $\pm$ 0.39 & 0.0526 $\pm$ 0.0013 & 12042 \\
        \bottomrule
    \end{tabular}
    }
\end{table*}

%\vspace{-0.2cm}

\FloatBarrier

\section{Conclusion}

%This project studied 3 Transformer-based approaches for particle reconstruction from sparse and unordered calorimeter hits: QueryDETR, DiffReco and FlowReco. The models use different reconstruction mechanisms while sharing the same task and evaluation procedure. Their results show that learned queries, diffusion and flow matching can all recover particle sets and estimate their energies and directions.

%DiffReco gave the lowest accepted-particle count error and the narrowest angular error distribution among the learned models. It also recovered the complete particle set most often across the displayed multiplicities. FlowReco remained competitive and achieved the narrowest accepted energy-residual distribution. All 3 models showed reduced energy precision when particles were close together, indicating that overlapping showers remain a challenging part of reconstruction. CLUEstering provided a clustering reference with broader distributions, although it uses different matching rules.

%These results support further investigation of modern machine-learning methods for calorimeter reconstruction. Future work can test their performance under a wider range of detector conditions and study hardware optimisation for possible trigger applications. The same approaches may also be useful in other reconstruction problems involving sparse, unordered measurements and an unknown number of objects.

This study evaluated three Transformer-based machine learning architectures for end-to-end particle reconstruction from calorimeter hits. Despite employing fundamentally different generative and query-based mechanisms, all three models are successfully able to reconstruct the particles. Our results demonstrate that learned queries, diffusion processes, and flow matching are all highly capable of mapping raw detector data into reconstructed particles while accurately estimating their kinematic properties outperforming the existing CLUEstering method in all evaluated benchmarks.

% PREVIOUS VERSION: Comparing the specific architectures, DiffReco achieved the most accurate particle multiplicity recovery and the narrowest angular error distribution, consistently identifying the complete particle set across varying multiplicity levels. FlowReco performed competitively alongside it, yielding the highest precision in energy estimation with the narrowest residual distribution. While resolving heavily overlapping showers remains inherently challenging—as evidenced by degraded energy precision at tight spatial separations across all models, the neural architectures significantly outperformed the CLUEstering baseline in both target recovery and resolution. The best throughput for the machine learning methods is achieved with FlowReco, which achieves only 13.8\% worse reconstruction speed than CLUEstering while keeping much high accuracy and performing the full particle identification eliminating the need for subsequent corrections and identification that is needed for CLUEstering. Ultimately, these findings suggest that end-to-end machine learning based reconstruction would be a suitable method for particle reconstruction for the HL-LHC era and beyond.

Comparing the specific architectures, QueryDETR achieved strong particle multiplicity recovery and the narrowest energy and angular error distributions. FlowReco performed competitively alongside it, achieving nearly identical particle recovery and similar energy precision with only two refinement steps. All three models obtained similar combined accuracy. Although resolving highly overlapping showers remains challenging, as shown by the poorer energy precision at small angular separations, the neural networks outperformed the evaluated CLUEstering baseline in both particle recovery and energy resolution. QueryDETR and FlowReco achieved similar inference throughput, processing approximately 12725 and 12042 events per second, respectively, compared with approximately 10145 for CLUEstering in the tested configurations. Ultimately, these findings suggest that end-to-end machine-learning-based reconstruction is a promising approach for the HL-LHC era and beyond, motivating further evaluation with larger datasets and more realistic detector conditions.

% The neural networks also predict particle identity directly within the reconstruction, however in this study this is limited to the calorimeter data only
%Ultimately, these findings strongly advocate for the integration of modern generative and continuous set-prediction models into high-granularity calorimeter reconstruction pipelines. Future work will explore the robustness of these models across a broader spectrum of simulated detector conditions and investigate hardware optimizations for potential low-latency trigger applications. Beyond the HL-LHC, these mathematical frameworks offer a versatile foundation for solving other complex physics challenges characterized by sparse, unorganized measurements and unknown object multiplicities.

\FloatBarrier

\section*{Acknowledgments}

% Lauri
Work done by Imperial College is funded by the Science and Technology Facilities Council (STFC) grant ST/W000636/1. We thank Max Marriott-Clarke for developing and providing the Geant4 simulation of the calorimeter used to produce the dataset in this work, and for helpful discussions.

\end{document}